\documentclass[prapplied,superscriptaddress,twocolumn]{revtex4}
\usepackage{amsmath,amssymb,graphicx,xcolor,bm,physics,soul}
\usepackage{todonotes,placeins}
\usepackage[normalem]{ulem}
\usepackage[T1]{fontenc}

\newcommand{\wvsout}[1]{}%{\wv{\sout{#1}}}
 	
\begin{document}

\title{Analogue Spin Simulators: How to keep the Amplitude Homogeneous}
%Quantum Photonic Spin Simulators that impose Homogeneous Amplitude

%\title{Quantum Photonic Spin Simulators Free of Bias}
%\title{Overcoming amplitude heterogeneity in quantum photonic spin solvers.}
\author{W. Verstraelen}

\affiliation{Division of Physics and Applied Physics, School of Physical and Mathematical Sciences, Nanyang Technological University, Singapore 637371, Singapore}

\author{P. Deuar}
\author{M. Matuszewski}

\affiliation{Institute of Physics, Polish Academy of Sciences, Al. Lotników 32/46, Warsaw PL-02-668, Poland}

\author{T. C. H. Liew}

\affiliation{Division of Physics and Applied Physics, School of Physical and Mathematical Sciences, Nanyang Technological University, Singapore 637371, Singapore}

\date{\today}

\begin{abstract}
A setup that simulates ground states of spin graphs would allow one to solve computationally hard optimisation problems efficiently. Current optical setups to this goal have difficulties decoupling the amplitude and phase degrees of freedom of each effective spin; risking to yield the mapping invalid, a problem known as amplitude heterogeneity. Here, we propose a setup with coupled active optical cavity modes, where this problem is eliminated through their particular geometric arrangement. Acting as an effective Monte Carlo solver, the ground state can be found exactly. By tuning a parameter, the setup solves XY or Ising problems.
\end{abstract}

\maketitle

\section{Introduction}

%\begin{itemize}
%\item %[Situation]
%\begin{itemize} 
    %\item 
    \subsection{Analogue spin simulators}
    
    Over the course of a few decades, the rise and spread of digital computers has transformed society completely. These rely mainly on Turing machines on a %the 
    von Neumann architecture, with continuous improvement until recently
    according to Moore's law. It has become clear however, that they are limited for certain tasks, leading to the conceptualisation of alternative setups. These include gate-based quantum computing \cite{Bruß_Leuchs_2018} and neural networks \cite{Rojas_1996} as well as analogue simulation (either classical or quantum \cite{Altman2021}).  
    %\item 
    Simulation refers to the ability of an accessible system to reproduce the physics of a less accessible one. While such a `less accessible system' may be studied for its own sake, it can also have applications. This is the case for simulating spin models on a graph. In fact, in the presence of frustration, these correspond to spin-glass systems \cite{Parisi2023} for which the determination of the ground state is an (NP- and QMA-) hard problem, needing a potentially exponential amount of time both in a classical or quantum setup \cite{Zeng_2019}. That also implies that simulating these spin models and finding their ground state would allow one to solve all other problems in these classes (such as travelling salesman or knapsack problems) with only polynomial overhead \cite{Stein_Newman_2013}. It has in fact been shown that all quantum computing tasks can be recast efficiently in instances of this problem through 
    so-called adiabatic quantum computing \cite{Aharanov2004}.

    There has been much interest to simulating such spin graphs especially with optics, because of their %their 
    ease of access and ability to operate away from equilibrium \cite{Mohseni2022,Kavokin2022,Stroev2023,Opala23}. Despite alternatives \cite{Kyriienko2019,Sigurdsson2019,Leonetti2021,Ohadi2017}, the by far most commonly proposed setup uses a mapping from optical phase to spin orientation. By default, phase in an oscillator has a freedom of $2\pi$, mapping straightforwardly to the orientation of an XY-spin \cite{Berloff2017,Struck2011,Nixon2013,Toebes2022}. Driving with degenerate parametric oscillators reduces the phase to only a few
    possible values (usually two), as in an Ising-spin. These are the %, as is done in so-called 
    Coherent Ising Machines and similar setups \cite{Mohseni2022,Marandi2014,Strinati2021PRAppl,Goto2016,Verstraelen2020,Piearangeli2019} or 
    Potts model solvers \cite{Kalinin2018PRL};  
    notably, some superconducting qubits are implemented similarly. Using time-multiplexing and electronic overheads, effective system sizes up to $10^4$ have been achieved \cite{Honjo2021}, while more recent schemes consider elimination of the electronics \cite{Strinati2021PRAppl}.

\begin{figure}[b]
\centering
\includegraphics[width=\linewidth]{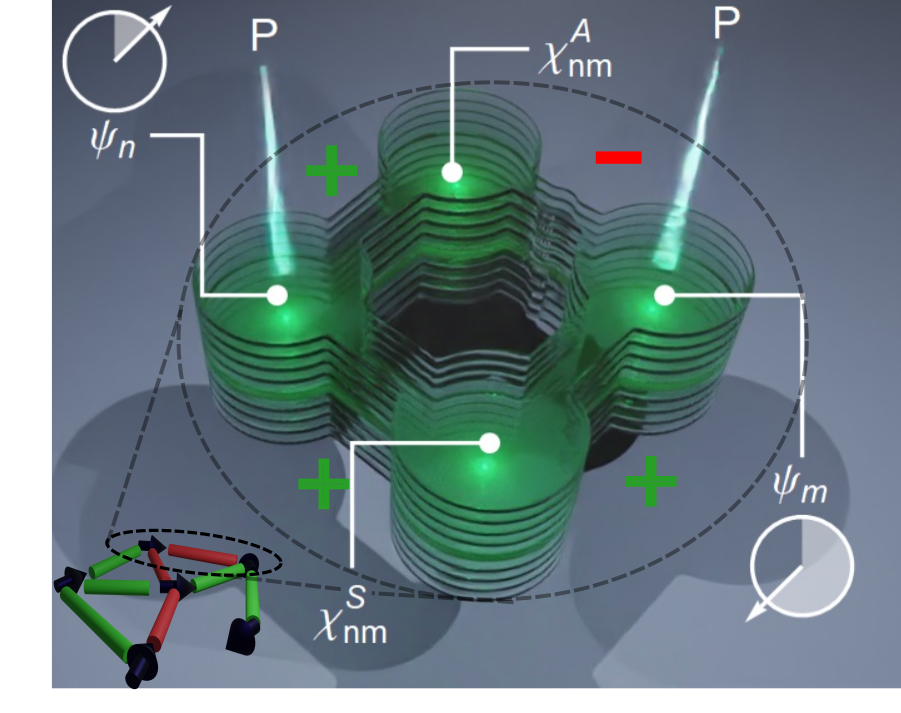}
%\missingfigure{}
%\includegraphics[width=4cm]{F1A.eps}
\caption{Depiction of the proposed scheme with polariton micropillars acting as coupled cavities. Two ``spin'' sites $\psi_n$ and $\psi_m$ are coherently coupled through the undriven ``connecting'' $\chi_{nm}^S$ and $\chi_{nm}^A$. The bump in the coupling between $\chi^A_{nm}$ and $\psi_m$ indicates its %the 
different path length giving this coupling a strength $-J$, while the others are all $+J$. Inset: $\psi_{n,m}$ correspond to a generic pair of connected spins in the graph, with either ferromagnetic or antiferromagnetic coupling.}
\label{fig:scheme}
\end{figure}

    In this work, we will introduce a novel flexible similator, depicted in Fig. \ref{fig:scheme}, that overcomes a critical limitation of many such previous setups.

\subsection{The problem: Amplitude heterogeneity}

  In the above
  systems, a linear -- but often dissipative -- direct coupling between the oscillators
  faithfully maps the spin model to  
  the coupling between on-site phases.

\begin{figure}
    \centering
    \includegraphics[width=\linewidth]{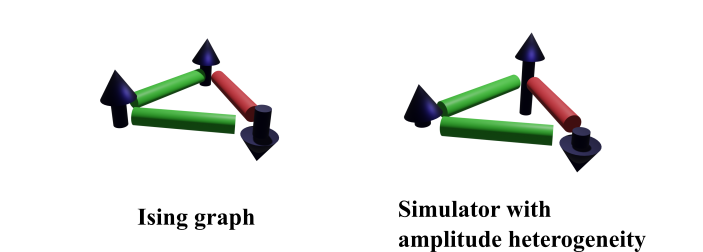}
    \caption{Amplitude heterogeneity. Consider an Ising model with $H=-\sum_{ij} J_{ij} s_i s_j$ on a triangular graph with two ferromagnetic ($J_{ij}=1$) and one antiferromagnetic ($J_{ij}=-1$) coupling, depicted in green (red), for variables $s=\pm 1$. It is frustrated: no Ising configuration can satisfy all couplings simultaneously. One of the degenerate ground states is shown. However, the mapping from spins $s_i$ to complex amplitudes $\psi_i$ is not exact, since the absolute values $\abs{\psi_i}$ are not normalized to unity. This leads to lower energy solutions which are unphysical in terms of the original Ising model. In this work, a scheme is introduced that avoids the occurrence of such amplitude heterogeneity.}
    \label{fig:heterogeneity}
\end{figure}

However, the analogue simulators typically have more degrees of freedom than the original spin model. To avoid affecting the dynamics, their value should remain fixed. In particular, for optical simulators, spin variables $s_i=e^{i\theta_i}$ are represented by a local order parameter $\psi_i=\abs{\psi_i}e^{i\theta_i}$, and consequently the Spin Hamiltonians 
  \begin{equation}\label{eq:SpinHam}
  H_\text{spin}=-\Re\sum_{\langle ij\rangle} \pm s_i^* s_j=-\sum_{\langle ij \rangle} \pm \cos{(\theta_i-\theta_j)}    
  \end{equation}
are mapped om 
\begin{equation}\label{eq:analogue}
 H_\text{analogue}=-\Re\sum_{\langle ij\rangle} \pm \psi_i^* \psi_j=-\sum_{\langle ij\rangle} \pm \abs{\psi_i}\abs{\psi_j} \cos{(\theta_i-\theta_j)},   
\end{equation} where $\langle nm \rangle$ are the edges of the graph and $\pm$ is positive(negative) for ferromagnetic(antiferromagnetic) couplings. Whereas in the XY-case the phase differences can have arbitrary values, in the Ising case, we have $\cos{(\theta_n-\theta_m)}=\pm 1$. see also the illustration in Fig \ref{fig:heterogeneity}).

When the amplitudes $\abs{\psi_i}$ are not kept fixed, it is easy to see that the spectra, and thus ground state configurations between \eqref{eq:SpinHam} anand \eqref{eq:analogue} will generally not match. This critical problem, ``Broken mapping due to amplitude heterogeneity'' \cite{Mohseni2022,Stroev2023} is known to ultimately invalidate the spin-solution \cite{Strinati2021PRL} and the simulation schemes. 

It is unsurprising, in a sense, that the computationally complex spin-glass problem cannot be equivalent to the completely linear problem that arises when the spins are replaced by harmonic oscillators such as noninteracting photons. 
Two mechanisms are known that can somewhat mitigate this problem.
The first one is to include optical nonlinearities, such as in exciton-polaritons \cite{Berloff2017}, so that a steady state is found by the nonlinear dynamics well above threshold \cite{Strinati2021PRL}. A clever choice of a sufficiently strong nonlinearity can be found, to have the strongest improvement, but it will generally be insufficient by itself to overcome heterogeneity \cite{Strinati2021PRL,Böhm2021}. The second mechanism is adapting the optical driving at every site continuously in a feedback loop to force a uniform intensity. This seems an intuitive fix, however, it is cumbersome on its own, as it introduces additional overhead, with potentially the need for an additional separation of timescales, and the necessity to estimate the threshold value \cite{Kalinin2018SciRep,Leleu2019}. Again, such a mechanism will give some moderate improvements in accuracy, but insufficient to overcome the bias from amplitude heterogeneity completely: even an exponentially slow operation will often fail to converge to the actual spin ground state \cite{Cummins2023}.

One may thus wonder if this amplitude heterogeneity is a fundamental issue of any optical spin solver, or if a scheme can be devised to bypass it directly.

\subsection{This work}
  
   Here, we will answer the latter question affirmatively and demonstrate the existence of such an analogue simulation scheme without amplitude heterogeneity.
    %\item 
   We show explicitly how it can be implemented in a system of coupled optical cavities experiencing gain and saturable absorption, such as those realizable 
   with exciton-polariton micropillars~\cite{Kim2011,Jacqmin2014,Winkler2016}.
  Such systems are already considered reasonable candidates for analogue
simulators given their fast operation time and accessibility~\cite{Amo2016,Kim2017} and are especially studied for XY spin models \cite{Kalinin2019,Kalinin2020,Kalinin2018SciRep,Kalinin2018NJP,Kalinin2018PRL}. % that is, accessible systems that can reproduce the physics of less accessible ones.   
    %\item 
     We consider a scheme where spin orientations are mapped to the phases of cavity fields. Its %The 
     unique feature is that the coupling between the spin-mapping cavities happens through a pair of intermediate cavities with coherent Josephson coupling in such a way %, as we will show,  PD -- this seems to break the flow 
     that the particle flow between effective spins is exactly cancelled -- as we will show -- and thus amplitude heterogeneity is avoided. Depending on a parameter, XY or Ising graphs are simulated. Our scheme works in a pulsed operation mode \cite{Feng2021,Chen2022}, as an effective Monte Carlo solver across different spin solutions; with a feedback system that ensures that the spin ground state solution survives. This paper is organised as follows: in \ref{sec:scheme} we introduce the scheme, and show how it can represent all spin configurations faithfully, i.e. with a homogeneous amplitude. In the next section \ref{sec:optimisation}, we then describe the approach to find the ground state of the spin model effectively. We conclude in section \ref{sec:conclusions}.
    %\item 

\section{A scheme that keeps the amplitude homogeneous \label{sec:scheme}}

It has been known that nonlinear oscillators with direct coupling generally tend to %do 
synchronize %in general 
according to various models \cite{WoutersPRB2008,Moroney2021,Kalinin2019}. However, it is believed that the resulting phase configuration matches a spin ground state only in the case of purely dissipative
couplings \cite{Kalinin2019}, a correlated loss process on neighbouring sites. It is clear that a lack of symmetry in such a setup generally affects the intensities inhomogeneously. Here, we explicitly look for an alternate setup, unaffected by these considerations.

Consider first two cavities with complex mean-field amplitudes, $\psi_1$ and $\psi_2$, which will represent the spins. Each cavity is coupled to two auxiliary `connecting' cavities with complex amplitudes, $\chi_{12}^S$ and $\chi_{12}^A$, (`Symmetric' and `Antisymmetric' for the effective coupling they will later induce), with the connectivity illustrated in Fig.~\ref{fig:scheme}.

We exploit the possibility to control the sign of the Josephson coupling between modes \cite{Ohadi2016}, but only need real (Hermitian) values, typically dominant in physical realizations \cite{Stepnicki2013}. The evolution of the system is described by complex Ginzburg-Landau equations \cite{Carusotto2013,Keeling2008} in the tight-binding limit~\cite{Stepnicki2013,Grelu2012}
\begin{align}\label{eq:GPE}
i\frac{\partial\psi_{1,2}}{\partial t}&=i\left(\frac{P-\gamma}{2}-\Gamma_\mathrm{NL}|\psi_{1,2}|^2\right)\psi_{1,2}-J\left(\chi_{12}^S\pm\chi_{12}^A\right)\nonumber\\
i\frac{\partial\chi_{12}^{S,A}}{\partial t}&=i\left(\frac{-\gamma}{2}-\Gamma_\text{NL}'\abs{\chi_{12}^{S,A}}^2\right)\chi_{12}^{S,A}
-J\left(\psi_1\pm\psi_2\right),%-\frac{i\gamma}{2}\chi_{S,A}-i\Gamma'_\mathrm{NL}|\chi_{S,A}|^2\chi_{S,A},
\end{align}
where the $\pm$ sign takes a $-$ value for the coupling between $\psi_2$ and $\chi_{12}^A$ only, and $+$ sign otherwise.
Here we consider the case of a local gain $P$ applied to the cavities $\psi_{1,2}$, a local decay $\gamma$ (dissipation) on all cavities,
and nonlinear losses (modelling gain saturation or saturable absorption into the system, and/or incoherent scattering and two-photon absorption out of the system) $\Gamma_\mathrm{NL}$ and $\Gamma'_\text{NL}$ applied to the two sets of modes, respectively. The choice of the coupling scheme (Fig.~\ref{fig:scheme}) ensures that, in the absence of the loss processes on the auxiliary modes, there is no effective coupling between the ``spin'' modes $\psi_1$ and $\psi_2$. This can be seen by considering (in the limit $\gamma=0$, $\Gamma'_\mathrm{NL}=0$):
\begin{equation}
i\left.\frac{\partial^2\psi_{1,2}}{\partial t^2}\right|_\text{coupling}=-J\left(\pdv{\chi_{12}^S}{t}\pm\pdv{\chi_{12}^A}{t} \right)=-2iJ^2\psi_{1,2},  
\end{equation}
That is,  $\psi_1$ and $\psi_2$ can evolve independently, allowing for solutions where $\abs{\psi_1}\equiv \abs{\psi_2}$.
Intuitively, transport from $\psi_1$ to $\psi_2$ is suppressed by destructive interference of the $\chi_1$ and $\chi_2$ paths. This prevents the occurrence of amplitude heterogeneity, and hence of broken mapping. In the presence of $\gamma\neq 0$, one finds that the stationary solutions of the system also fix $|\psi_1|^2=|\psi_2|^2$. Specifically:
\begin{align}
\chi_{12}^{S,A}&=\frac{2Ji}{\gamma}\left(\psi_1\pm\psi_2\right) \label{eq:chiXY}\\ 
I_0:=\abs{\psi_1}^2&=\abs{\psi_2}^2=\frac{P-\gamma}{2\Gamma_\text{NL}}-\frac{4J^2}{\gamma\Gamma_\text{NL}}\label{eq:I0orig}
\end{align}
%
% %
% \begin{align}
% \chi_{1,2}=-iJ\frac{\left(\psi_1\pm\psi_2\right)}{\Gamma}\label{eq:chistationary}
% \end{align}
%
%yields $|\psi_1|^2=|\psi_2|^2=\left(W-2J^2/\Gamma\right)/\Gamma_\mathrm{NL}$. 
Note that the phases of $\psi_{1,2}$ can still be freely chosen.

\subsection{XY Graphs} We now consider a graph network. At its nodes, we consider `spin' sites $\psi_n$ and the edges of the graph are implemented by the pair of `connecting' cavities $\chi_{nm}^{S/A}$ with the %same 
connectivity scheme from Fig. \ref{fig:scheme}. For generality, we will consider arbitrarily connected graphs. In practice, by coupling neighboring cavities one may only be able to access planar graphs, however every other, potentially all-to-all connected graph 
can be mapped onto these \cite{Garey1976,Cuevas2016}. 
In addition, it is in principle possible to couple spatially remote cavities via external optics \cite{Kalinin2020} or a mask \cite{Nixon2013}.

For a `spin' site $\psi_m$ connected to $n_\text{nb}^m$ neigbouring spin sites, %it is immediately clear that 
\eqref{eq:I0orig} generalizes to $I_0^{(m)}=(P_m-\gamma)/(2\Gamma_\text{NL})-4 n_\text{nb}^m J^2/(\gamma\Gamma_\text{NL})$. In general, this value would not be constant for homogenous $P_m=P$,
as $n_\text{nb}^m$ varies across main sites $m$ in a generic graph. However, in order to avoid any bias, we require $I_0^{(m)}=I_0$. We can ensure this condition in different ways. The first approach comprises of extending the original spin graph with `extra' %additional 
dangling (single neighbour) 
spins that do not affect the ground state in the subgraph corresponding to the original configuration; and implementing the scheme based on this extended graph. 

\begin{figure}[h]
\centering%\missingfigure{} 
\includegraphics[width=\columnwidth]{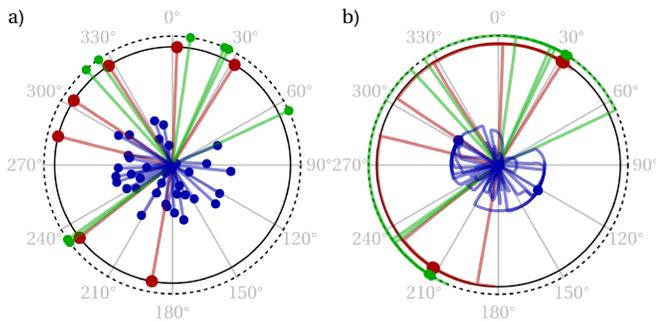}
%{amplitudeandphasetogether.eps}
\caption{Absolute amplitude (radial component) and phase (angles) for a random graph composed of seven `spin' sites (red), eight `extra' sites (green) that ensure the main sites have all the same number of neighbours (four for the considered example), and `connecting' sites (blue). Time evolution is shown with a linearly increasing opacity, up to a stationary state denoted by the spots. (a) XY type behaviour where the `spin' sites adopt arbitrary phases; (b) Ising type behaviour where the `spin' site phases become binary. The black circle corresponds to the analytically predicted magnitude $\sqrt{I_0}$.
Parameters: $J/\Gamma_\mathrm{NL}=0.5$; $P/\Gamma_\mathrm{NL},=12+2$; $\gamma/\Gamma_\mathrm{NL}=4$; and  $\Gamma'_\mathrm{NL}/\Gamma_\mathrm{NL}=0$ in the XY-case (a) or Ising case (b). The graph has a 50\% connectivity, all antiferromagnetic to ensure frustration to make it nontrivial. Even though the `extra' sites are used, we have added contribution +2 to $P/\Gamma_\mathrm{NL}$ that compensates exactly for the losses $8n_\text{NB} J^2/\gamma$ through the neighbouring sites.}
\label{fig:SpinModelBehaviour}
\end{figure}

Figure~\ref{fig:SpinModelBehaviour}a shows a polar plot of the resulting state of a typical graph, obtained by solving numerically from an initial low intensity random complex noise state and time evolving toward a stationary state. In this calculation the nonlinear loss on the `connecting' sites is set to zero ($\Gamma'_\mathrm{NL}=0$). We find that the `main' sites (red) rapidly grow in intensity and equilibrate at the analytically expected value of $I_0$, while preserving their phase from the initial noise. The `extra' sites (green) show {a }similar behaviour, although they settle at slightly higher intensity as they have only one connected neighbour each. The `connecting' sites (blue) settle with a variety of different amplitudes and phases. %Focusing on 
The `spin' sites  %rearranged - PD
correspond effectively to a valid configuration of the corresponding XY graph, since %as 
they each have $\mathcal{U}(1)$ freedom in the phases, while their intensities are the same.  
Similar behaviour to Fig.~\ref{fig:SpinModelBehaviour}a was observed for other %different 
random graphs, of different sizes and %different 
choices of initial noise.

The second  approach to ensure homegeneous intensities is by setting $P_m=P+8n_\text{NB}^m J^2/\gamma$ at each site. We highlight that such a non-uniform pump scheme is manifestly different from the ones considered in \cite{Kalinin2018SciRep,Leleu2019} as 
in our case the proper values of $P_m$ are a priori known.

\subsection{Ising Graphs} Upon introducing a
nonlinear loss on the `connecting' sites ($\Gamma'_\mathrm{NL}>0$), a further symmetry breaking appears, that is: the behaviour illustrated in Fig.~\ref{fig:SpinModelBehaviour}b emerges. The system still attains a stationary state, however,
after approaching the intensity $I_0$ the `spin' sites coalesce their phases to one of two values, corresponding to $\mathbb{Z}_2$ freedom at each site. That is, the system develops into an effective Ising state. In Appendix \ref{ap:gammanl}, we show how these Ising configurations remain fixed points of the nonlinear dynamics in presence of such ($\Gamma'_\mathrm{NL}>0$).

\section{Optimising spin models \label{sec:optimisation}}

\subsection{Extracting the spin-model energy}

\begin{figure*}
    \centering
    \includegraphics[width=\linewidth]{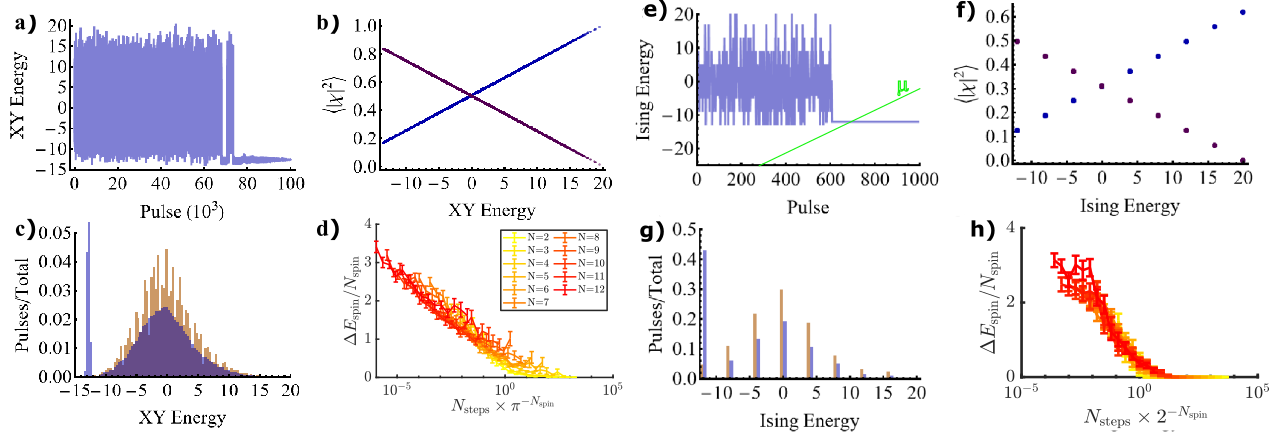}
    \caption{Option 1 (all in one picture) Solving XY (a-d) and Ising (e-h) graphs. Panels (a-c)[(e-g)] depict the solution of the same random XY [Ising] graph and parameters (with $P_r$ taking the value of $P$)
    of \eqref{fig:SpinModelBehaviour}. (a)[(e)] The spin energy stabilizes when the feedback strength $P_f\propto \mu(j)$ becomes sufficiently strong. An example of a rising $\mu(j)$ is schematically added. (b)[(f)] as analytically predicted, $E_\text{spin}$ is directly proportional to $\ev{\abs{\chi}^2}$, where $\ev{\cdot}$ denotes the spatial average, selecting either $\chi^S$ or $\chi^A$ out of every coupling--allowing the scheme to work. (c)[(g)] Compares %shows the comparison of 
the distribution of XY[Ising] energies $E_\text{spin}$ from the photonic simulator across pulses $j$ (blue) with the raw density of states (as obtained froma brute force search) (orange). These results show that the ground state solution is exactly found, and as the simulator remains in this configuration, it forms a peak which keeps on growing indefinitely. This behaviour was verified for numerous individual graphs (d)[(e)] Statistics in the performance of different random graphs at multiple sizes, in finite time. Depicted is the Excess energy above the true ground state $\Delta E_\text{spin}$ as function of a rescaled number of pulses (yellow-red for different graph sizes $N_\text{sites} =2-12$, each datapoint is the average over 30 graphs). Each curve shows that $\Delta E_\text{spin}$ and hence the distance from the true ground state becomes vanishingly small for sufficient pulses. The collapse for different sizes demonstrates the scaling behaviour that allows one to extrapolate that the scheme would apply to graphs of arbitrary size. For these graphs, 50\% all-to-all connectivity was chosen, which in turn has 50\% Ferromagnetic or Antiferromagnetic coupling, reflecting realistic large graph problems. To compensate for external losses, in the XY case (d) extra dangling sites were used in combination with offset in $P_r,P_f$, while in the Ising case (h) only the offset was used without extra dangling sites. Results of the Ising case are less noisy since the spin states remain gapped (arbitrary perturbations wouldn't lead to other configurations). All pulses used a total duration of 1000$\Gamma_\mathrm{NL}^{-1}$}
    \label{fig:resultsXYandIsing}
\end{figure*}

% \begin{figure}
%     \includegraphics[width=\linewidth]{XYplusIsingclassical_2.png}
%     \caption{Option 2 (first part)}
% \end{figure}

% \begin{figure}
%     \centering
%     \includegraphics[width=\linewidth]{XYplusIsingclassical_2bis.png}
%     \caption{Option 2 (second part)}
%     \label{fig:enter-label}
% \end{figure}

We have discussed how a random spin configuration of either an XY or Ising type Hamiltonian can form in the coupled cavity system. Considering the example of exciton-polaritons, the states form on a timescale of picoseconds (or even shorter in some materials) and hence by using a pulsed operation scheme, it is possible to sample a few trillion states in seconds in subsequent pulses $j=1\ldots N_\text{pulses}$.
However, so far the discussed scheme does not distinguish these many states.

Let us recap that for each pair of spins $(n,m)$ connected in the original graphs, the corresponding `spin' sites $\psi_n$ and $\psi_m$ are coupled through a pair of `connecting' sites $\chi_{nm}^S$ and $\chi_{nm}^A$.
In the limit $\Gamma'_\text{NL}=0$, giving XY model states, the stationary intensities are:
\begin{equation}
|\chi_{nm}^{S,A}|^2=\frac{8J^2}{\gamma^2}I_0\left(1\pm\cos\left(\phi_n-\phi_m\right)\right)\label{eq:chi2},
\end{equation}
where the $+$ sign applies to the $\chi_{nm}^S$ and $-$ to $\chi_{nm}^A$ (The Ising case involves a simple rescaling, as shown in Appendix \ref{ap:gammanl}).

% The energy functional of the underlying spin model is:

% \begin{equation}\label{eq:spinE}
%     E_\text{spin}=-\sum_{<nm>} \pm \cos{(\phi_n-\phi_m)},
% \end{equation}

Defining $E_\text{spin}$ by
\eqref{eq:SpinHam}, we immediately see that 

\begin{equation}
E_\text{spin}\propto -\sum_{\langle nm \rangle}\abs{\chi_{nm}^X}^2 + C,
\end{equation}
where $X=S,A$ for ferromagnetic or antiferromagnetic coupling respectively and $C$ is an irrelevant constant offset. 
The important feature is that the configuration that minimises $E_\text{spin}$ also maximises $\abs{\chi_{nm}^X}^2$, a quantity that can be easily accessed in experiment by summing emission intensity from specific `connecting' sites.

\subsection{Reaching the ground state}

Having completed the mapping, we now turn towards the protocol of finding the ground state, which has lowest $E_\text{spin}$, or, equivalently, highest $\sum_{<nm>}\abs{\chi_{nm}^X}^2.$

We will work in a pulsed setup, and thus define a periodic sequence of operations $j=1\ldots N_\text{pulses}$. Each period $j$ contains two stages, a `readout' stage and a `feedback' stage. We also define a real function $\mu(j)$, that increases linearly (or at least monotonically) with pulse number $j$.

\paragraph{readout stage} During the first half of the period, the system evolves according to (\ref{eq:GPE})
with a fixed `spin' site gain $P=P_r$, during which a stationary state spin state is formed, such as the ones in Fig. \ref{fig:SpinModelBehaviour}. Then, also the `coupling' site intensities
\eqref{eq:chi2} are extracted to obtain $\sum_{<nm>}\abs{\chi_{nm}^X}^2$.

\paragraph{feedback stage} During the second half of the period, one pumps the `spin' sites with a gain
\begin{equation}\label{eq:classfb}
P_f := \mu(j)\sum_{<nm>}\abs{\chi_{nm}^X}^2,
\end{equation}
everyhing else remaining the same. Unlike $P_r$ in the readout state, $P_f$ in the feedback stage is thus different for each pulse. Based on this value of $P_f$ for the pulse, two situations can occur.

As long as $P_f$ is sufficiently small, the condensate vanishes and the next pulse $j+1$ forms a new one with different phases. This means that the system stochastically samples different spin states many times across the pulses $j$. By choosing $\mu(0)$ to be sufficiently low, we can always ensure that this situation occurs for the initial pulses. 

Now, we have considered the parameter $\mu(j)$ to rise slowly with pulse count $j$. This will lead at some point for $P_f$ to be sufficiently large for the second situation to occur. In this, the condensate is maintained at substantial intensity, large enough for the spin configuration to survive until the next pulse $j+1$. 

If $\mu(j)$ is increased slowly enough, this is guaranteed to happen first when $E_\text{spin}$ is minimal, as per the definition of \eqref{eq:classfb}. Furthermore, the fact that $\mu(j)$ is monotonically increasing, ensures that the survival of the spin configuration corresponding to this pulse continues throughout all subsequent pulses.

In Figure \ref{fig:resultsXYandIsing}, we show how this picture is confirmed numerically for both the 
XY and Ising models. The first three panels for each model show the effect of applying the procedure above to an individual spin graph, as
simulated by Eq.~\eqref{eq:GPE}.
The effect of losing phase coherence below threshold is included by adding a small inhomogeneous random noise at the end of each feedback phase. As desired, after a certain number of pulses the system settles to an effective low energy state due to the feedback scheme. This is only possible because of the correlation between the intensities of the S and A
sets of `connecting' sites with the effective spin energy $E_\text{spin}$, which is shown in panels (b,f). The obtained effective energy distributions are plotted in panels (c,g) and compared to the distribution attained with a brute force search. While the photonic simulator initially samples across the same distribution, there emerges a sharp peak at the optimum (lowest) energy, which would grow in size if more pulses in the sequence were considered. 

Now that we have illustrated the workings of the mechanism for an individual graph, we proceed 
to verify that it generalizes well to arbitrary graphs of different sizes (Panels (d,h)). We tested graph sizes of $N=2-12$ spins, with 30 random variations each. For the individual graph sizes, we see a steady decay of the excess energy above the true ground state to zero, meaning that all distance metrics to the true ground state must also decay, and highlighting that the ground stand can be exactly obtained with a sufficient amount of pulses. While the number of pulses to achieve a certain accuracy increases exponentially with system size (as expected for an NP-hard problem) good collapse of the rescaled results emerges. This allows us to extrapolate the validity to the (thermodynamic) limit of arbitrarily large spin-glass graphs.

\section{Conclusions \label{sec:conclusions}}

Analogue spin solvers hold great prospect as an emerging technology to solve complex computational problems efficiently. Despite a large amount of work in this area, there has been a major limitation of this principle, in that the optics-to-spin mapping becomes invalid due to heterogeneity of the optical amplitudes.
Here, we have introduced a novel scheme with coupled optical resonators (e.g. cavities with gain media or exciton-polaritons). Depending on a parameter, this scheme is able so solve both XY or Ising spin problems. Most importantly, the scheme keeps the intensities completely homogeneous and is thus devoid of the aforementioned amplitude heterogeneity. This makes it possible for arbitrarily large spin-graph problems to be solved without bias.

\begin{acknowledgments}
We gratefully acknowledge interesting discussions with N.G. Berloff, O. Kyriienko, M. Wouters and M. Weitz. WV and TCHL were supported by the Singaporean Ministry Education, via the Tier 3 grant MOE2018-T3-1-002 and Tier 2 grant MOE-T2EP50121-0020. MM and PD acknowledge %s
support from Polish National Science Center grant %s 
2021/43/B/ST3/00752. % and 2018/31/B/ST2/01871, respectively.
\end{acknowledgments}

%%%%%%%%%%%%%%%%%%%%%%%%%%%%%%%%%%%%%%%
%Bibliography
%%%%%%%%%%%%%%%%%%%%%%%%%%%%%%%%%%%%%%%

%\bibliographystyle{plain}

\appendix
\section{Role of $\Gamma_{nl}'$ and the Ising case \label{ap:gammanl}}

To understand  better how the choice of $\Gamma'_\mathrm{NL}$ determines XY or Ising type behaviour, let us again consider the plaquette of two `spin' cavities $\psi_{1,2}$ representing spins coupled in parallel by two `connecting' cavities $\chi^{S/A}$. We write $\psi_1=\sqrt{n_1}e^{i\phi_1}$ and consider the evolution of the phase, yielding:
\begin{equation}
\frac{\partial\psi_1}{\partial t}+\frac{\partial\psi_1^*}{\partial t}=\frac{\cos\phi_1}{\sqrt{n_1}}\frac{\partial n_1}{\partial t}+2\sqrt{n_1}\frac{\partial\cos\phi_1}{\partial t}\label{eq:phaseevolution1}
\end{equation}
The terms in Eq.~(\ref{eq:phaseevolution1}) can be evaluated as:
\begin{align}
\frac{\partial\psi_1}{\partial t}+\frac{\partial\psi_1^*}{\partial t}&=\left(P-\gamma-2n_1\Gamma_\mathrm{NL}\right)\sqrt{n_1}\cos\phi_1\notag\\
&\hspace{5mm}+iJ\left(\chi^S+\chi^A-\chi^{S*}-\chi_2^{A*}\right)\\
\frac{\partial n_1}{\partial t}&=\left(P-\gamma-2n_1\Gamma_\mathrm{NL}\right)n_1\notag\\ \label{eq:Isingdn}
&\hspace{5mm}+iJ\left(\chi^S+\chi^A\right)\psi_1^*-iJ\left(\chi^{S*}+\chi^{A*}\right)\psi_1
\end{align}
Equation~(\ref{eq:phaseevolution1}) can then be written as:
\begin{align}
iJ\left(\chi^S+\chi^A-\chi^{S*}-\chi^{A*}\right)-iJ\frac{\cos\phi_1}{\sqrt{n_1}}\left(\chi^S+\chi^A\right)\psi_1^*\notag\\
+iJ\frac{\cos\phi_1}{\sqrt{n_1}}\left(\chi^{S*}+\chi^{A*}\right)\psi_1=2\sqrt{n_1}\frac{\partial\cos\phi_1}{\partial t}\label{eq:phaseevolution2}
\end{align}
Let us assume that $\chi^{S,A}$ follows its stationary value, which is given by a similar equation to Eq.~\eqref{eq:chiXY}, but here we allow for $\Gamma'_\mathrm{NL}\neq0$:
\begin{align}\label{eq:Isingchi}
\chi^{S,A}=\frac{2iJ\left(\psi_1\pm\psi_2\right)}{\gamma+2\Gamma'_\mathrm{NL}|\chi_{S,A}|^2}
\end{align}
Equation~(\ref{eq:phaseevolution2}) becomes:
\begin{align}
-2J^2\frac{\psi_1+\psi_2}{\gamma+2\Gamma'_\mathrm{NL}|\chi^S|^2}-2J^2\frac{\psi_1-\psi_2}{\gamma+2\Gamma'_\mathrm{NL}|\chi_A|^2}\notag\\
-2J^2\frac{\psi_1^*+\psi_2^*}{\gamma+2\Gamma'_\mathrm{NL}|\chi^S|^2}-2J^2\frac{\psi_1^*-\psi_2^*}{\gamma+2\Gamma'_\mathrm{NL}|\chi^A|^2}\notag\\
+\frac{2J^2\cos\phi_1}{\sqrt{n}_1}\left(\frac{\psi_1+\psi_2}{{\gamma+2\Gamma'_\mathrm{NL}|\chi^S|^2}}+\frac{\psi_1-\psi_2}{{\gamma+2\Gamma'_\mathrm{NL}|\chi^A|^2}}\right)\psi_1^*\notag\\
+\frac{2J^2\cos\phi_1}{\sqrt{n}_1}\left(\frac{\psi_1^*+\psi_2^*}{{\gamma+2\Gamma'_\mathrm{NL}|\chi^S|^2}}+\frac{\psi_1^*-\psi_2^*}{{\gamma+2\Gamma'_\mathrm{NL}|\chi^A|^2}}\right)\psi_1\notag\\
=2\sqrt{n_1}\frac{\partial\cos\phi_1}{\partial t}\label{eq:phaseevolution3}
\end{align}
In the case of $\Gamma'_\mathrm{NL}=0$ all the denominators are equal, the $\psi_2$ terms cancel, and the left-hand side of the equation is zero. In this case the phase $\phi_1$ becomes constant and a stationary state is allowed without needing a specific value of $\phi_1$. Similar arguments hold for $\phi_2$, which can take any phase independent of $\phi_1$, giving XY type behaviour.

In the case of $\Gamma'_\mathrm{NL}\neq0$ the denominators in Eq.~(\ref{eq:phaseevolution3}) are not generally equal and we have a finite value of $\partial\cos\phi_1/\partial t$, implying that the phase continues evolving. However, in the special cases $\psi_1=\pm\psi_2$, it is readily seen that the terms on the left-hand side of Eq.~\ref{eq:phaseevolution3} again vanish. Consequently, we find that the system can only reach a stable stationary state when $\psi_1=\pm\psi_2$. Thus, we obtain an Ising type behaviour. We note that this solution implies that one of the $\chi^{S,A}$ of each pair reaches a finite intensity $\abs{\chi^X}^2$ with a phase difference $\pi/2$ while the other is empty. As in the XY case, this solution prevents any polariton current from $\psi_1$ to $\psi_2$, allowing to keep the amplitude homogeneous again. By inserting \eqref{eq:Isingchi} in the Gross-Pitaevskii equation (\eqref{eq:GPE}), we obtain the algebraic equation 

\begin{equation}
\Gamma_\text{NL}'\abs{\chi^X}^3+\frac{\gamma}{2}\abs{\chi^X}-2J\sqrt{\frac{P-\gamma}{2\Gamma_\text{NL}}}=0.    
\end{equation}

Solving for $\abs{\chi^X}$, we can correct $I_0$ once more, in the case of $n_\text{nb}$ neighbours this becomes
\begin{equation}    
I_{0,\text{mod}}:=\abs{\psi_n}^2=\frac{P-\gamma}{2\Gamma_\text{NL}}-\frac{4 n_\text{nb} J^2}{(\gamma+2\abs{\chi^X}^2\Gamma_\text{NL}' )\Gamma_\text{NL}}.
\end{equation}

One way to ensure that this quantity remains homogeneous is by adapting the incoherent pump $P$ to
\begin{equation}
 P\rightarrow P_n=P+8n_\text{nb} J^2/\left(\gamma+2\abs{\chi^X}^2\Gamma_\text{NL}'\right)
 \end{equation}

Note that we have some additional numerical evidence (not presented) that a similar scheme also applies with a coherent nonlinearity in the `connecting' sites instead of the nonlinear losses ($\Gamma'_{NL}\rightarrow i \alpha'$).

\end{document}